\documentstyle[prl,aps,eqsecnum,epsfig,multicol]{revtex}

\def\Journal#1#2#3#4{{#1}{\bf #2}, #3 (#4)}

\def\NIMA{{Nucl. Instrum. Methods}~{\bf A}}
\def\NPA{{Nucl. Phys.}~{\bf A}}

\def\PLB{{Phys. Lett.}~{\bf B}}
\def\PLC{Phys. Repts.\ }
\def\PRL{Phys. Rev. Lett.\ }
\def\PRD{{Phys. Rev.}~{\bf D}}
\def\PRC{{Phys. Rev.}~{\bf C}}
\def\ZPC{{Z. Phys.}~{\bf C}}

\begin{document}
\draft

\title{Event-by-event fluctuations in Mean $p_T$ and Mean $e_T$ \\
in $\sqrt{s_{NN}}=130~$GeV Au+Au Collisions}

\author{
K.~Adcox,$^{40}$
S.{\,}S.~Adler,$^{3}$
N.{\,}N.~Ajitanand,$^{27}$
Y.~Akiba,$^{14}$
J.~Alexander,$^{27}$
L.~Aphecetche,$^{34}$
Y.~Arai,$^{14}$
S.{\,}
% ...
\\(PHENIX Collaboration)\\
}
\address{
% ...
$^{3}$Brookhaven National Laboratory, Upton, NY 11973-5000, USA\\
% ...
$^{14}$KEK, High Energy Accelerator Research Organization, Tsukuba-shi, Ibaraki-ken 305-0801, Japan\\
% ...
$^{27}$Chemistry Department, State University of New York - Stony Brook, Stony Brook, NY 11794, USA\\
% ...
$^{34}$SUBATECH (Ecole des Mines de Nantes, IN2P3/CNRS, Universite de Nantes) BP 20722 - 44307, Nantes-cedex 3, France\\
% ...
$^{40}$Vanderbilt University, Nashville, TN 37235, USA\\
% ...
}
\maketitle

\begin{abstract}
Distributions of event-by-event fluctuations of the mean transverse 
momentum and mean transverse energy near mid-rapidity have been
measured in Au+Au collisions at $\sqrt{s_{NN}}=130~$GeV
at RHIC. By comparing the distributions to what is expected 
for statistically independent particle emission, the magnitude of 
non-statistical fluctuations in mean transverse momentum is determined 
to be consistent with zero. Also, no significant non-random 
fluctuations in mean transverse energy are observed.  By constructing
a fluctuation model with two event classes that preserve the mean and
variance of the semi-inclusive $p_T$ or $e_T$ spectra,
we exclude a region of fluctuations in $\sqrt{s_{NN}}=130~$GeV 
Au+Au collisions.

\end{abstract}


\begin{multicols}{2}   
\narrowtext            

\section{Introduction}

Phase instabilities near the QCD phase transition can result in non-statistical
fluctuations that are detectable in final state observables~\cite{Hei01}. 
These instabilities, which may occur due to random color fluctuations
\cite{Mro93}, critical behavior at the QCD tri-critical point~\cite{Ste98},
or fluctuations from the decay of a Polyakov loop condensate~\cite{Dum01},
can result in a broadening of the transverse momentum or transverse energy
distributions of produced particles for different classes of events. 
This phenomenon is expected to be detected experimentally by searching for 
deviations of the distributions of the event-by-event mean transverse 
momentum $M_{p_T}$ or mean transverse energy $M_{e_T}$ of produced 
particles from the random distributions expected for statistically 
independent particle emission.

An event-by-event analysis of $M_{p_T}$ was 
previously performed for $158~A~$GeV/c Pb+Pb Collisions at the CERN
SPS by Experiment NA49~\cite{App99}. In that analysis, the $M_{p_T}$
distributions measured over the rapidity range $4<y_{\pi}<5.5$ and $p_T$ range 
$0.005<p_T<1.5~$GeV/c, were found to be consistent with random fluctuations.
NA49 also performed an event-by-event analysis of the $K/\pi$ ratio
\cite{Afa01}, showing only very small deviations from random fluctuations.
With an increase of $\sqrt{s_{NN}}$ to $130~$GeV in RHIC collisions, 
unprecedented energy densities have been observed~\cite{ppg001}, hence
conditions may be more favorable for a phase transition from hadronic matter 
to a Quark-Gluon Plasma which may be indicated in non-random fluctuations.
Presented here is an event-by-event analysis of $M_{p_T}$ fluctuations 
and the first measurement of $M_{e_T}$ fluctuations at mid-rapidity at RHIC.

\section{Analysis}

The PHENIX experiment~\cite{Mor98} consists of four
spectrometers designed to measure simultaneously hadrons, leptons,
and photons produced in nucleus-nucleus, proton-nucleus, and proton-proton
collisions at RHIC. The two central arm spectrometers, which are located
within a focusing magnetic field and each cover $\pm 0.35$ in pseudorapidity
and $\Delta \phi = 90^{\circ}$ in azimuthal angle, are utilized in 
this analysis. The primary interaction trigger was defined using the 
Beam-Beam Counters (BBC)~\cite{Ike98} and Zero Degree Calorimeters 
(ZDC)~\cite{Adl01}.  Events are selected with a 
requirement that the collision vertex along the beam axis has $|z|<20~$cm
as measured by both the BBC and ZDC. Event centrality is defined using 
correlations in the BBC and ZDC analog response as described in~\cite{ppg001}.
For the present analysis, the events are classified according to the $0-5\%$, 
$0-10\%$, $10-20\%$, and $20-30\%$ most central events.

The drift chamber~\cite{Ria98} is used in conjunction with the innermost
pad chamber, called PC1, to measure the transverse momentum of charged 
particles traversing the PHENIX acceptance. A fiducial section of the drift 
chamber is chosen to minimize the effect of time-dependent variations in the 
performance of the detector during the data-taking period. The fiducial 
volume of the $M_{p_T}$ analysis spans an azimuthal range of 
$\Delta\phi = 58.5^{\circ}$ and covers the pseudorapidity
range $|\eta| < 0.35$. Reconstructed tracks~\cite{Mit01} are required to 
contain a match to a hit in PC1 to insure that the tracks are well 
reconstructed in three dimensions for reliable momentum determination. 

The $M_{e_T}$ distribution is determined from clusters reconstructed in
the two instrumented sectors of the Lead-Scintillator electromagnetic calorimeter
\cite{Mor98,Dav98,ppg002}.  The quantity $e_T$ is defined as the transverse energy 
per reconstructed calorimeter cluster as described in~\cite{ppg002}, which can 
include clusters that have been merged. The effects of cluster merging on the 
$M_{e_T}$ distribution are discussed later. The fiducial volume of the $M_{e_T}$ 
analysis spans an azimuthal range of $\Delta \phi = 45^{\circ}$ and covers $|\eta|<0.35$.

There are no acceptance nor efficiency corrections applied to the semi-inclusive 
$p_T$ or $e_T$ distributions prior to the calculation of $M_{p_T}$ or $M_{e_T}$.
These corrections do not vary from event to event and are identical for data and 
mixed events; therefore they do not modify the values of the fluctuation quantities 
defined later. The $M_{p_T}$ distributions are calculated using the formula
\begin{equation}
M_{p_T} = (1/N_{tracks})\sum^{N_{tracks}}_{i=1}{p_T}_i,
\end{equation}
where $N_{tracks}$ is the number of tracks in the event that pass the cuts
outlined above and lie within the $p_T$ range $0.2<p_T<1.5~$GeV/c.
Similarly, the $M_{e_T}$ distributions are calculated using the formula
\begin{equation}
M_{e_T} = (1/N_{clus})\sum^{N_{clus}}_{i=1}{e_T}_i,
\end{equation}
where $N_{clus}$ is the number of calorimeter clusters in the event that lie within
the $e_T$ range $0.225<e_T<2.0~$GeV. An event is excluded from the analysis 
if $N_{tracks}$ or $N_{clus}$ is below a minimum value to insure 
that there are a sufficient number of tracks or clusters to determine a mean
and to exclude background events.  This minimum 
value for the $0-5\%$, $0-10\%$, $10-20\%$, and $20-30\%$ centrality classes, 
respectively, is 40, 30, 20, and 10 for the $M_{p_T}$ analysis and 30, 20, 10, 
and 10 for the $M_{e_T}$ analysis.  Table I lists statistics pertaining to the 
data samples used to determine $M_{p_T}$ and Table II lists the statistics 
pertaining to the data samples used to determine $M_{e_T}$. The events used 
for the $M_{p_T}$ and $M_{e_T}$ analyses are considered independently of each other. 

In order to compare the $M_{p_T}$ and $M_{e_T}$ distributions to what is expected 
for statistically independent particle emission, mixed events are considered 
as the baseline for the random distribution. To obtain a precision comparison, it is 
important to match the number of tracks or clusters along with the mean of the
semi-inclusive distribution of the mixed events to that of the data.
Therefore, in both analyses, mixed events are constructed by pre-determining
the number of charged particle tracks or calorimeter clusters in the mixed event
$N_{mix}$ by directly sampling the corresponding data $N_{tracks}$ or
$N_{clus}$ distributions. Figure 1 shows a comparison of the $N_{tracks}$ distributions
from the data and the normalized mixed event $N_{mix}$ distribution for the $0-10\%$ 
centrality class.  Once $N_{mix}$ is determined, a mixed event is filled with 
$p_T$ or $e_T$ values from the data with the following criteria: a) no two $p_T$ 
or $e_T$ values from the same data event are allowed to reside in the same 
mixed event, b) only $p_T$ or $e_T$ values passing all cuts in the determination 
of $M_{p_T}$ or $M_{e_T}$ from the data events are placed in a mixed event, 
and c) only data events from the same centrality class are used to construct a 
mixed event corresponding to that class.  Once a mixed event is filled with 
$N_{mix}$ tracks or clusters, its $M_{p_T}$ or $M_{e_T}$ is calculated in 
the same manner as for the data events.  

For both analyses, the data contain a fraction of tracks or clusters within close 
physical proximity that have merged into a single track or cluster.  This fraction
is estimated by embedding simulated single-particle events that are processed 
through a detailed simulation of the detector response into real data events, which
are then reconstructed in the same manner as the data.  For the $0-5\%$ centrality
class, we estimate that 6\% of the tracks and 5\% of the clusters are affected.

For the $M_{p_T}$ analysis, tracks that are merged into a single reconstructed
track typically have similar values of $p_T$.  The result is a slightly lower value 
of $N_{tracks}$ which causes a slight broadening in the width of the $M_{p_T}$ 
distribution due to the reduced statistics per event.  However, since the 
$N_{tracks}$ data distribution is directly sampled during the construction of 
mixed events, the effect of merged tracks cancels for comparisons between the 
data and mixed events.

For the $M_{e_T}$ analysis, the effect of merged clusters is complicated by the fact
that a single cluster is reconstructed with an $e_T$ corresponding to the sum of the
two (or more) particles contributing to the cluster.  To understand this effect 
on the mixed events, we note that the fraction of merged clusters within a 
data event increases with event multiplicity.  Also, many of the data events 
with the lowest $M_{e_T}$ coincide with the lowest multiplicity events since they 
contain few, if any, merged clusters that would yield a higher $M_{e_T}$.  
When the merged clusters in the data events are randomly 
redistributed among the mixed events, low multiplicity mixed events can contain more 
merged clusters than the data events with the same multiplicity, resulting in a gross 
upward shift in $M_{e_T}$ for those mixed events.  This results in apparent excess 
non-random fluctuations at low $M_{e_T}$.  Conversely, high multiplicity mixed events 
can contain fewer merged clusters than the data events with the same multiplicity, 
resulting in a gross downward shift in $M_{e_T}$ for those mixed events.  However, 
since the mean is taken over more clusters in this case, the effective shift in 
$M_{e_T}$ is reduced at high $M_{e_T}$, and the apparent non-random fluctuations 
are much less pronounced.  An estimate of the magnitude of this effect is presented 
later.

\section{Results}

To compare directly the semi-inclusive $p_T$ distribution to the
$M_{p_T}$ distribution assuming statistically independent particle emission,
the closed form prescription outlined in~\cite{Tan01} is used.  This prescription
describes the semi-inclusive $p_T$ distribution using a Gamma distribution,
\begin{equation}
  f(p_T) = f_{\Gamma}(p_T,p,b) = \frac{b}{\Gamma(p)}(b p_T)^{p-1} e^{-b p_T},
\end{equation}
where $p$ and $b$ are free parameters that are related to the mean and standard
deviation of the semi-inclusive distribution as
\begin{equation}
  p = \frac{<p_T>^2}{\sigma_{p_T}^2}, \; \; b = \frac{<p_T>}{\sigma_{p_T}^2},
\end{equation}
where
\begin{equation}
   \sigma_{p_T} = (<p_T^2> - <p_T>^2)^{1/2}.
\end{equation}
The reciprocal of $b$ is the inverse slope parameter of the $p_T$ distribution.
With the track multiplicity distribution assumed to be a negative binomial 
distribution, \mbox{$f_{NBD}(N_{tracks},1/k,<N_{tracks}>)$}, the $M_{p_T}$ distribution 
can be calculated using
\begin{equation}
  g(p_T)=\sum^{N_{max}}_{N=N_{min}}{f_{NBD}(N,1/k,<N>) f_{\Gamma}(p_T,Np,Nb)},
\end{equation}
where the loop is over $N_{tracks}$ from $N_{min}$ to $N_{max}$, which are the limits of 
the multiplicity.
The value of the negative binomial distribution parameter $k$ is given by
\begin{equation}
   \frac{1}{k} = \frac{\sigma_{p_T}^2}{<N_{tracks}>^2} - \frac{1}{<N_{tracks}>}.
\end{equation}
Therefore, given $<p_T>$, $\sigma_{p_T}$, and $<N_{tracks}>$ extracted from the
semi-inclusive $p_T$ distribution, the expected random $M_{p_T}$ distribution
can be calculated.  Figure 2 shows the 
$M_{p_T}$ distribution for the $0-5\%$ centrality class. Overlayed on the data 
as a dotted curve is the result of the calculation. The agreement between the 
data distribution and the calculation illustrates the absence of large 
non-statistical fluctuations in the data. The remainder of this article will 
quantify the amount of non-statistical fluctuations observed and place limits 
on the level of fluctuations that can be present in central Au+Au collisions 
at $\sqrt{s_{NN}}=130~$GeV.

To quantify the magnitude of the deviation of fluctuations from the expectation of
statistically independent particle emission, the magnitude of the 
fluctuation $\omega_T$ in the transverse quantity $M_T$, representing 
$M_{p_T}$ or $M_{e_T}$, is defined as
\begin{equation}
   \omega_T = \frac{(<M_T^2> - <M_T>^2)^{1/2}}{<M_T>} = \frac{\sigma_{M_T}}{<M_T>}.
\end{equation}
The value of $\omega_T$ is calculated independently for the data distribution
and for the baseline, or mixed event, distribution.
The difference in the fluctuation from a random baseline distribution
is defined as
\begin{equation}
   d = \omega_{(T,~data)} - \omega_{(T,~baseline)}.
\end{equation}
The sign of $d$ is positive if the data distribution contains a correlation, such
as Bose-Einstein correlations~\cite{Wie99}, when compared to the baseline distribution.
The fraction of fluctuations which deviate from the expectation of statistically 
independent particle emission is given by
\begin{eqnarray}
   F_T & = \frac{(\omega_{(T,~data)}-\omega_{(T,~baseline)})}{\omega_{(T,~baseline)}}\nonumber \\
       & = \frac{(\sigma_{(T,~data)}-\sigma_{(T,~baseline)})}{\sigma_{(T,~baseline)}},
\end{eqnarray}
where $\sigma_{(T,~data)}$ refers to the standard deviation of the event-by-event
$M_T$ data distribution and $\sigma_{(T,~baseline)}$ is the corresponding quantity
for the baseline, or mixed event, distribution.
In the absence of a common language for the analysis of $M_{p_T}$ and $M_{e_T}$
fluctuations, the commonly used fluctuation quantity $\phi_{T}$~\cite{Gaz92} is also 
presented in order to compare this measurement to previous results~\cite{App99}.
The quantity $d$ is related directly to $\phi_{T}$ via
\begin{eqnarray}
   \phi_{T} & = (\sigma_{(T,~data)} - \sigma_{(T,~baseline)}) \sqrt{<N_T>}\nonumber \\
	& = d <M_T> \sqrt{<N_T>},
\end{eqnarray}
where $N_T$ represents $N_{tracks}$ or $N_{clus}$.
The quantity $\phi_{T}$ is related to $F_T$ by
\begin{equation}
   \phi_{T} = F_T \sigma_{(T,~baseline)} \sqrt{<N_T>}.
\end{equation}
The standard deviation of the semi-inclusive spectra can be approximated by
\mbox{$\sigma_{(T,incl.)} \approx \sigma_{(T,baseline)} \sqrt{<N_T>}$} \cite{Tan01}, where 
$\sigma_{(T,incl.)}$ is the standard deviation of the semi-inclusive distribution 
as defined in Eq. 3.3.  Therefore, $\phi_{T}$ is simply the fraction of non-random 
fluctuations in the event-by-event mean $p_T$ or $e_T$, $F_T$, scaled by 
$\sigma_{(T,incl.)}$.  An advantage of $F_T$ over $\phi_{T}$ is that measurements 
expressed in $F_T$ can be directly compared without further scaling.

The magnitudes of any non-random fluctuations are established by comparing the
data distributions to the mixed event distributions, which serve as the
random baseline distributions. For this purpose, the mixed event distributions
are normalized to minimize the $\chi^2$ value with respect to the data
distributions.  Figure 3 and Figure 5 show the $M_{p_T}$ and $M_{e_T}$ 
distributions for all four centrality classes (data points) with the corresponding 
mixed event $M_{p_T}$ and $M_{e_T}$ distributions overlayed on the data as dotted 
curves. The broadening of the distributions for less central collisions are due to 
the reduction in $<N_{tracks}>$ or $<N_{clus}>$. Shown in Figure 4 and Figure 6 are the
residuals between the data and mixed events, defined for each bin $i$ as 
\mbox{$residual_i = (M_{(T, data)_i} - M_{(T, mixed)_i})/\sigma_i$}, in units of standard
deviations, for each centrality class.  The shapes of the residual distributions
are primarily driven by the normalization procedure applied to the mixed events.

For the $M_{p_T}$ distributions, the data and mixed event distributions are indistinguishable.
However, the upper $M_{e_T}$ edges of the data and mixed event $M_{e_T}$ distributions show 
good agreement while the lower $M_{e_T}$ edge of the data distributions are slightly wider 
than the mixed event distribution. If this low $e_T$ effect were physical, it would imply 
fluctuations with slightly more low $e_T$ photons since the effect is not seen in the 
$M_{p_T}$ distribution for charged particle tracks.  However, some of the excess fluctuations 
at low $e_T$ can be attributed to the effects of cluster merging previously discussed.
The magnitude of this effect has been investigated using a Monte Carlo simulation which 
calculates $M_{e_T}$ after reproducing the calorimeter cluster separation distribution, 
the $N_{clus}$ distribution, and the semi-inclusive $e_T$ distributions from the data.  
The fluctuations in the $M_{e_T}$ distribution with this effect included in each event are 
compared to a simulated mixed event $M_{e_T}$ distribution constructed from the same 
generated dataset using the same procedure that is applied to the data.  In this manner, 
it is estimated that the cluster merging effect contributes an additional 
$F_T$ = 1.5\%, 2.1\%, 0.9\%, and less than 0.01\% to the non-random fluctuations for the 
$0-5\%$, $0-10\%$, $10-20\%$, and $20-30\%$ centrality classes, respectively. The simulation 
confirms that the cluster merging effect significantly contributes only to the lower $M_{e_T}$ 
edge of the distribution.  The remainder of the excess low $e_T$ fluctuations 
is likely due to correlated low energy background. GEANT~\cite{Bru94} simulations 
indicate that the primary background contribution is produced by low energy electrons 
and muons that scatter off of the pole tips of the central arm spectrometer magnet 
but still pass the cluster selection cuts. Because of the difficulty in quantifying 
the contribution of background to the excess fluctuations, the present $M_{e_T}$ 
data are taken to indicate an upper limit on non-statistical fluctuations rather 
than an indication of true non-statistical fluctuations.

The values of $\omega_{T}$, $d$, $F_T$, and $\phi_T$ for each centrality class
using the mixed events as the random baseline distribution are tabulated in 
Table III for $M_{p_T}$ and Table IV for $M_{e_T}$.  The errors quoted for 
these quantities include statistical errors and systematic errors due to 
time-dependent variations over the data-taking period. The systematic errors 
are estimated by dividing each dataset into nine subsets with each subset 
containing roughly equal numbers of events.  For the $M_{p_T}$ analysis, the
systematic errors contribute to 81\%, 88\%, 76\%, and 75\% of the total error in
$\omega_{T}$ and 85\%, 88\%, 80\%, and 85\% of the total error in the variables
$d$, $F_T$, and $\phi_t$ for the $0-5\%$, $0-10\%$, $10-20\%$, and $20-30\%$ centrality
classes, respectively.  The corresponding values for the $M_{e_T}$ analysis are
a 67\%, 63\%, 81\%, and 82\% contribution to the total errors in $\omega_T$,
and a 64\%, 63\%, 81\%, and 82\% contribution to the total errors in $d$, $F_T$,
and $\phi_t$ for each centrality class.  The cluster merging contribution
estimates noted above are not applied to the values quoted in Table IV.

\section{Discussion}

Based upon the fluctuation measurements presented here, certain fluctuation 
scenarios in RHIC Au+Au collisions at $\sqrt{s_{NN}}=130~$GeV are excluded.
For this purpose, we consider two variations of a model that contains two 
classes of events with a difference of effective temperature, defined as
\mbox{$\Delta T = T_2 - T_1$}, where $T_2$ is the inverse slope parameter of the event
class with the higher effective temperature, and $T_1$ is the inverse slope
parameter of the event class with the lower effective temperature. The
first variation, {\it Model A}, will consider a case where the means of the
semi-inclusive $p_T$ spectra for the two event classes are identical, but
the standard deviations are different.  The second variation, {\it Model B}, will
consider a case where the means of the semi-inclusive $p_T$ spectra are
different, but the standard deviations are identical.  Since the semi-inclusive
$p_T$ distribution is an observed quantity, the two event classes must be
constrained in such a way that the mean and standard deviation of the
final semi-inclusive $p_T$ distribution remains constant while the 
effect of the fluctuation manifests itself in the $M_{p_T}$ distribution.

The dual event class model is applied to the determination of the 
sensitivity to fluctuations in $M_{p_T}$ for the $0-5\%$ centrality class as 
follows. Returning to the prescription outlined in~\cite{Tan01}, the 
semi-inclusive transverse $p_T$ spectrum can be parameterized by the 
$f_{\Gamma}(p_T,p,b)$ distribution defined in Eq. 3.1.  For both model variations, 
the fraction of events in the event class with the higher effective 
temperature is defined as
\begin{equation}
   q = \frac{(N_{events})_{class~1}}{(N_{events})_{class~1}+(N_{events})_{class~2}}.
\end{equation}
The $p_T$ distribution of the combined sample can then be expressed as
\begin{equation}
	f(p_T) = q \Gamma(p_T,p_1,b_1) + (1-q) \Gamma(p_T,p_2,b_2),
\end{equation}
where \mbox{$T_1=1/b_1$} and \mbox{$T_2=1/b_2$}.

For Model A, the semi-inclusive $p_T$ distributions of each event class 
are constrained to have the same mean, so we require
\begin{equation}
	\mu = p/b = p_1/b_1 = p_2/b_2.
\end{equation}
The variance of the final semi-inclusive $p_T$ distribution 
for Model A is constrained by
\begin{equation}
	\frac{\sigma^2}{\mu^2} = \frac{1}{p} = \frac{q}{p_1} + \frac{(1-q)}{p_2}.
\end{equation}
With these constraints, the choice of a value for $q$ and the effective temperature
of one event class is sufficient to extract the remaining parameters from which
sensitivity estimates for fluctuations in $M_{p_T}$ are obtained.

For Model B, the semi-inclusive $p_T$ distributions of each event class are allowed 
to have different means, $\mu_1$ and $\mu_2$, so the mean of the total semi-inclusive
distribution can be expressed as \mbox{$\mu = q\mu_1 + (1-q)\mu_2$}.  Defining a mean shift,
$\Delta\mu$, as $\Delta\mu = \mu_2 - \mu_1$, we obtain
\begin{equation}
	\mu_2 = \mu + q\Delta\mu.
\end{equation}
Allowing $p_1$ = $p_2$ and applying the constraint that the variances of the two event
classes are identical, yields
\begin{equation}
\frac{1}{p_1} = \frac{\frac{1}{p} - 
q(1-q)(\frac{\Delta\mu}{\mu})^2}{1+q(1-q)(\frac{\Delta\mu}{\mu})^2}.
\end{equation}
With a choice of values for $q$ and $\Delta\mu$, the remaining parameters can
be calculated, including $\Delta T$.

Both variations of the dual event class model are implemented in a Monte Carlo 
simulation in the following manner.  The number of particles in an event is 
determined by sampling the $N_{tracks}$ data distribution, approximated by
a Gaussian distribution fit to the data. The $p_T$ of each particle 
in an event is determined individually by sampling the appropriate 
$\Gamma(p_T,p,b)$ distribution fit to the semi-inclusive $p_T$ data distribution,
which yields $p = 0.8$ and $b = 2.46$ for $0-5\%$ centrality.  The $p_T$ of each
particle is restricted to the $p_T$ range of the measurement. With $N_{tracks}$
and the $p_T$ distribution determined, the $M_{p_T}$ for a given number of events
is calculated.  The generated $M_{p_T}$ distribution with $q=0$ for either model
variation is found to be statistically consistent with the mixed event $M_{p_T}$ 
distribution.

The data contain a fraction of background particles that did not originate from 
the collision vertex that effectively dilute the sensitivity to non-random 
fluctuations. To address this, a fraction of the particles in an event are 
randomly tagged as background particles, whose $p_T$ distribution is
then generated with a separate parameterization prior to calculating $M_{p_T}$
for an event.  The level of background contamination is estimated by processing 
HIJING~\cite{Wan91} Au+Au events through a software chain that includes a 
detailed GEANT simulation~\cite{Bru94} with the complete PHENIX detector 
geometry included, followed by a detailed simulation of the detector electronics 
response~\cite{Mit01}, whose output is then processed by track, cluster, and 
momentum reconstruction using the identical software and input parameters as 
is used for the data analysis. It is estimated that 11\% of the tracks and 
26\% of the clusters are due to background particles, independent of centrality 
class over the centrality range of these measurements.  The estimated $p_T$ 
and $e_T$ distributions for the background particles are well parameterized 
by exponential distributions. Again, the majority of the $e_T$ background occurs 
at low $e_T$, so any correlated background would most likely contribute to the 
lower side of the $M_{e_T}$ distribution.

To determine the sensitivity to fluctuations within the dual event class model,
the fluctuation fraction, $q$, and the value of $p_1$ for Model A and $\Delta\mu$
for Model B are varied and the $M_{p_T}$ distribution is generated at each step. 
A chi-square test is then performed on the generated $M_{p_T}$ 
distribution with respect to the mixed event data $M_{p_T}$ distribution. For a 
given value of $q$, the $\chi^2$ result increases as $\Delta T$ increases, which 
allows a fluctuation exclusion region to be defined for the single degree of
freedom.  The curves in Figure 7 show the lower exclusion boundaries for the 
$0-5\%$ centrality $M_{p_T}$ measurement at the 95\% Confidence Level as a 
function of $q$ and $\Delta T$ for both variations of the model.  If the sensitivity
is determined based upon the non-mixed data distribution, the lower exclusion
boundary increases by less than 2 MeV for all values of $q$ for either model.
Also, for all values of $q$ in either model, the estimated background contribution 
degrades the sensitivity estimates by $\Delta T$ = 3 MeV for both models.

A recent model of event-by-event fluctuations where the temperature parameter
$T = 1/b$ fluctuates with a standard deviation $\sigma_T$ on an event-by-event
basis~\cite{Kor01}, can be simply related to $F_T$:
\begin{equation}
   \frac{\sigma_T^2}{<T>^2} = \frac{2~F_T}{p(<n>-1)},
\end{equation}
where $p=0.8$ is the semi-inclusive parameter extracted from the present
data. For the $0-5\%$ centrality class, the present measurement establishes a 
95\% Confidence Limit of \mbox{2.6 $\times 10^{-3}$} for 
\mbox{$\sigma_T^2 / <T>^2$}, or 5\% for \mbox{$\sigma_T / <T>$}.

\section{Conclusions}

The fluctuations in the $M_{p_T}$ distributions for all centrality classes
are consistent with the presence of no fluctuations in excess of the random
expectation. The magnitude of $F_T$ in all cases is positive, which may be
due to the presence of Hanbury-Brown-Twiss correlations. The fluctuations in 
the $M_{e_T}$ distributions do have a small non-statistical component, much
of which is attributable to the effects of merged clusters, the
remainder of which are taken to indicate an upper limit on non-statistical
fluctuations in transverse energy. By defining a dual event class model, limits 
are set on the amount of $M_{p_T}$ fluctuations that can be present in 
the angular aperture $|\eta| < 0.35$ and $\Delta\phi = 58.5^{\circ}$ in
$\sqrt{s_{NN}}=130~$GeV Au+Au collisions. During the RHIC run of 2001, PHENIX 
has taken data for $\sqrt{s_{NN}}=200~$GeV Au+Au collisions with about a factor 
of four increase in azimuthal angular acceptance for both the $M_{p_T}$ and 
$M_{e_T}$ analyses, which will allow the measurements to be extended toward
more peripheral collisions.

\section{Acknowledgements}
 

We thank the staff of the RHIC Project, Collider-Accelerator, and Physics
Departments at Brookhaven National Laboratory and the staff of the other
PHENIX participating institutions for their vital contributions.
We acknowledge support from the Department of Energy, National Science
Foundation, and Dean of the College of Arts and Sciences, Vanderbilt
University (U.S.A), Ministry of Education, Culture, Sports, Science, and
Technology and the Japan Society for the Promotion of Science (Japan),
Russian Academy of Science, Ministry of Atomic Energy of Russian
Federation, Ministry of Industry, Science, and Technologies of Russian
Federation (Russia), Bundesministerium fuer Bildung und Forschung,
Deutscher Akademischer Auslandsdienst, and Alexander von Humboldt Stiftung
(Germany), VR and the Wallenberg Foundation (Sweden), MIST and the Natural
Sciences and Engineering Research Council (Canada), Conselho Nacional de
Desenvolvimento Cient\'{\i}fico e Tecnol\'ogico and Funda\c c\~ao de
Amparo \`a Pesquisa do Estado de S\~ao Paulo (Brazil), Natural Science
Foundation of China (People's Republic of China), IN2P3/CNRS (France),
Department of Atomic Energy and Department of Science and Technology
(India), Korea Research Foundation and Center for High Energy Physics
(Korea), the U.S. Civilian Research and Development Foundation for the
Independent States of the Former Soviet Union, and the US-Israel
Binational Science Foundation.





\vspace{-0.25cm}
\begin{figure}
\centerline{\epsfig{file=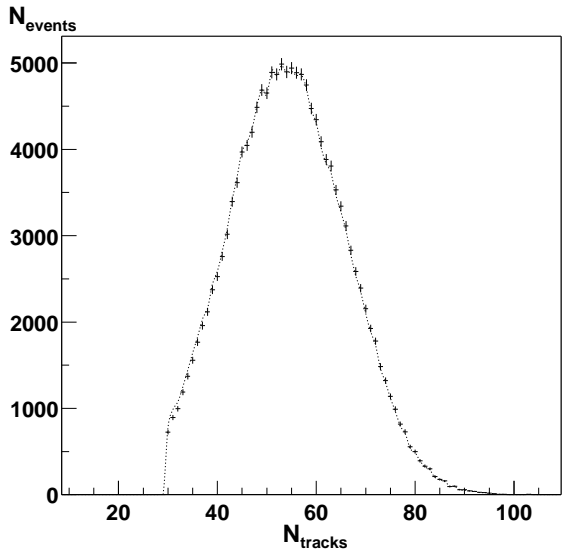,width=1.0\linewidth}}
\caption[]{The $N_{tracks}$ distribution for the $0-10\%$ centrality
class (data points) compared to the $N_{mix}$ distribution
from the mixed event sample (curve).}
\label{fig:1}
\end{figure}
   
\vspace{-0.25cm}
\begin{figure}
\centerline{\epsfig{file=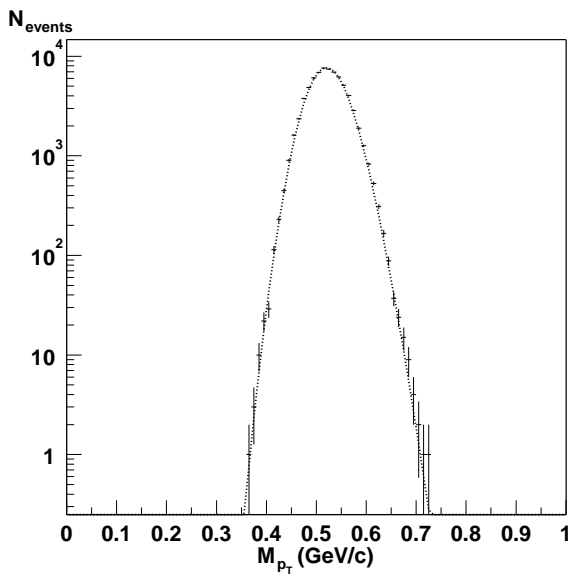,width=1.0\linewidth}}
\caption[]{The $M_{p_T}$ distribution for the $0-5\%$ centrality class.
The curve is the result of a $\Gamma$ distribution calculation
with parameters taken from the semi-inclusive $p_T$ spectra.}
\label{fig:2}
\end{figure}
   
\vspace{-0.25cm}
\begin{figure}
\centerline{\epsfig{file=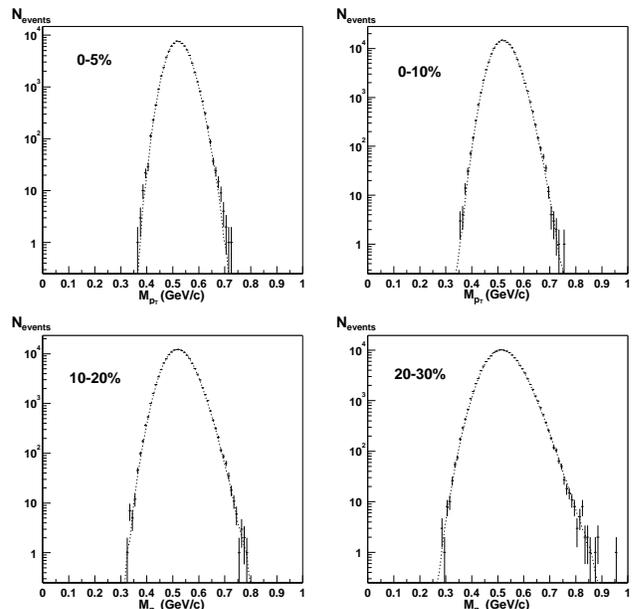,width=1.0\linewidth}}
\caption[]{The $M_{p_T}$ distributions for all centrality classes.
The curves are the random baseline mixed event distributions.}
\label{fig:3}
\end{figure}

\vspace{-0.25cm}
\begin{figure}
\centerline{\epsfig{file=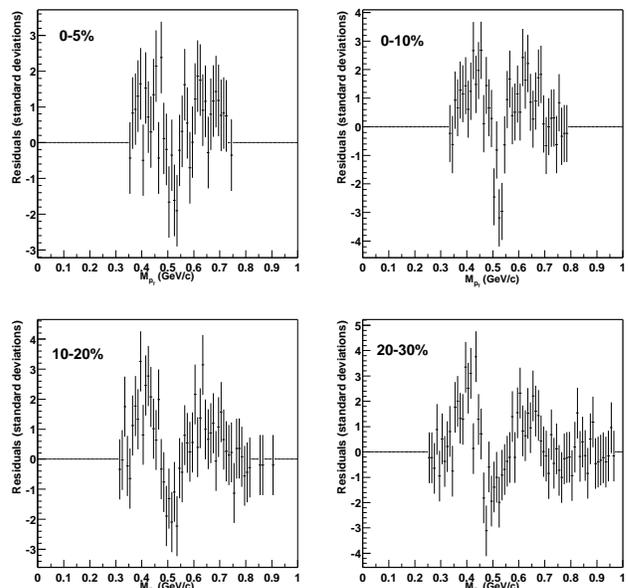,width=1.0\linewidth}}
\caption[]{The residual distribution between the data and mixed
event $M_{p_T}$ spectra as a function of $M_{p_T}$ in units of
standard deviations for all centrality classes.  The total $\chi^2$ 
and the number of degrees of freedom for the $0-5\%$, $0-10\%$, $10-20\%$, 
and $20-30\%$ centrality classes are $89.0/37$, $155.7/40$, $163.3/47$, 
and $218.4/61$, respectively.}
\label{fig:4}
\end{figure}


\vspace{-0.25cm}
\begin{figure}
\centerline{\epsfig{file=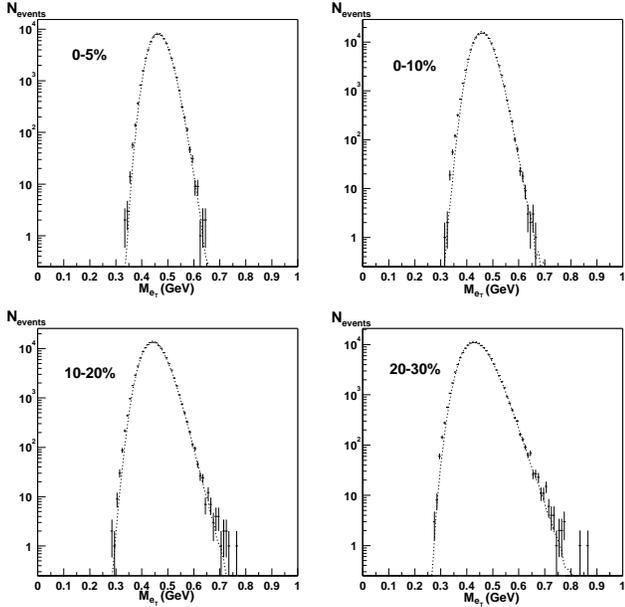,width=1.0\linewidth}}
\caption[]{The $M_{e_T}$ distributions for all centrality classes.
The curves are the random baseline mixed event distributions. The source
of differences in the data and mixed event distributions are addressed
in the text.}
\label{fig:5}
\end{figure}

\vspace{-0.25cm}
\begin{figure}
\centerline{\epsfig{file=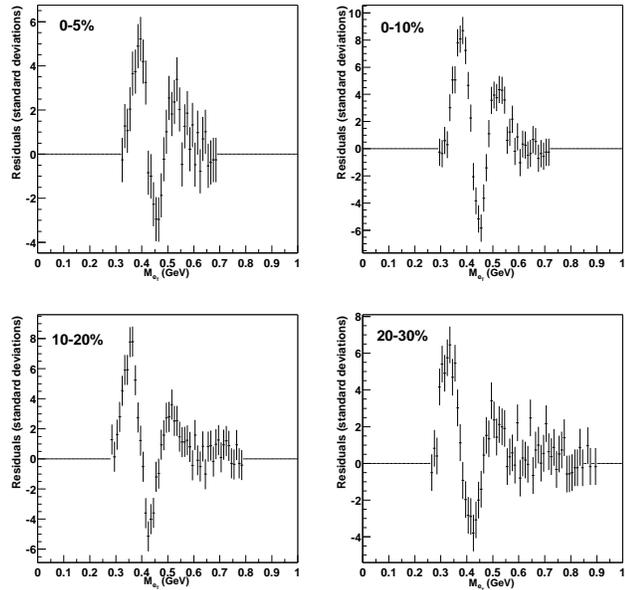,width=1.0\linewidth}}
\caption[]{The residual distribution between the data and mixed
event $M_{e_T}$ spectra as a function of $M_{e_T}$ in units of
standard deviations for all centrality classes.  The total $\chi^2$ 
and the number of degrees of freedom for the $0-5\%$, $0-10\%$, $10-20\%$, 
and $20-30\%$ centrality classes are $310.0/32$, $896.4/36$, $678.7/47$, 
and $553.9/53$, respectively.  A large fraction of the residual contributions are 
due to the effects of cluster merging.}
\label{fig:6}
\end{figure}

\vspace{-0.25cm}
\begin{figure}
\centerline{\epsfig{file=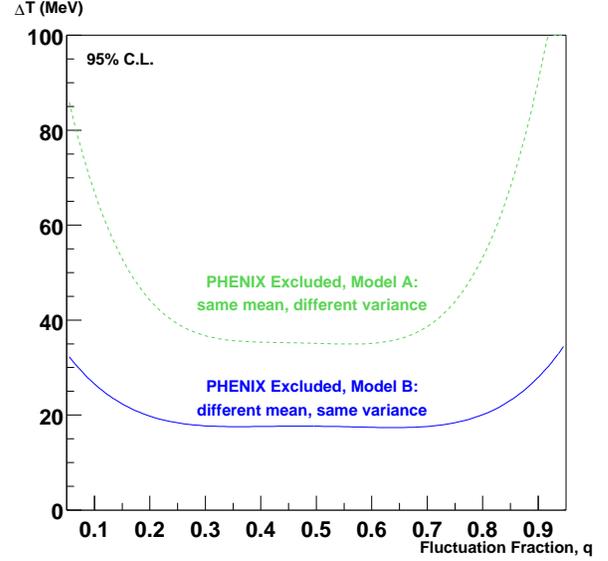,width=1.0\linewidth}}
\caption[]{The PHENIX sensitivity to non-random fluctuations in the 
two variations of the dual event class model that are excluded at the 
95\% confidence level by the $M_{p_T}$ analysis in the $0-5\%$ centrality class. 
The fraction of events, $q$, in the class of events with the lower inverse 
slope parameter (event class 1) is plotted on the horizontal axis while 
the difference in inverse slope parameter between event class 1 
and event class 2, $\Delta T$, is plotted on the vertical axis.
The curves represent the lower boundaries of the excluded regions.}
\label{fig:7}
\end{figure}

\newpage   

\begin{table} 
\label{tab:ptStats}
\caption{Statistics pertaining to the $M_{p_T}$ analysis. The values
of $<M_{p_T}>$ are quoted for $0.2 < p_T < 1.5~GeV/c$ and
are not corrected for efficiency or acceptance.}
\begin{tabular}[]{c|cccc}
Centrality&
$0-5\%$&
$0-10\%$&
$10-20\%$&
$20-30\%$ \\ \hline
\multicolumn{5}{c}{Data}\\ \hline
$N_{events}$& 72692 & 149236 & 149725 & 150365 \\
$<N_{tracks}>$& 59.6 & 53.9 & 36.6 & 25.0 \\
$\sigma_{N_{tracks}}$& 10.8 & 12.2 & 10.2 & 7.8 \\
$<M_{p_T}>~(MeV/c)$& 523 & 523 & 523 & 520 \\
$\sigma_{p_T}~(MeV/c)$& 290 & 290 & 290 & 289 \\
$\sigma_{M_{p_T}}~(MeV/c)$& 38.6 & 41.1 & 49.8 & 61.1 \\ \hline
\multicolumn{5}{c}{Mixed Events}\\ \hline
$<M_{p_T}>~(MeV/c)$& 523 & 523 & 523 & 520 \\
$\sigma_{M_{p_T}}~(MeV/c)$& 37.8 & 40.3 & 48.8 & 60.0 \\
\end{tabular}
\end{table}

\vspace{1cm}
\begin{table} 
\label{tab:etStats}
\caption{Statistics pertaining to the $M_{e_T}$ analysis. The values
of $<M_{e_T}>$ are quoted for $0.225 < e_T < 2.0~GeV$ and are not corrected
for efficiency or acceptance.}
\begin{tabular}[]{c|cccc}
Centrality&
$0-5\%$&
$0-10\%$&
$10-20\%$&
$20-30\%$ \\ \hline
\multicolumn{5}{c}{Data}\\ \hline
$N_{events}$&
69224  & 138882 & 140461 & 137867 \\
$<N_{clus}>$& 68.6 & 62.1 & 41.6 & 28.0 \\
$\sigma_{N_{clus}}$& 11.6 & 13.2 & 10.8 & 8.3 \\
$<M_{e_T}>~(MeV)$& 466 & 462 & 448 & 439 \\
$\sigma_{e_T}~(MeV)$& 267 & 265 & 258 & 253 \\
$\sigma_{M_{e_T}}~(MeV)$& 34.1 & 36.2 & 43.0 & 51.8 \\ \hline
\multicolumn{5}{c}{Mixed Events}\\ \hline
$<M_{e_T}>~(MeV)$& 466 & 462 & 448 & 439 \\
$\sigma_{M_{e_T}}~(MeV)$& 32.7 & 34.4 & 41.3 & 50.0 \\
\end{tabular}
\end{table}


\vspace{1cm}
\begin{table} 
\label{tab:ptFluc}
\caption{Fluctuation quantities for the $M_{p_T}$ analysis.}
\begin{tabular}[]{c|cccc}
Centrality&
$0-5\%$& $0-10\%$& $10-20\%$& $20-30\%$ \\ \hline
$\omega_{(T,data)}$(\%)& 
$7.37 \pm 0.10$& 
$7.85 \pm 0.13$&
$9.52 \pm 0.14$&
$11.7 \pm 0.21$ \\
$d(\%)$&
$0.14 \pm 0.15$& 
$0.16 \pm 0.19$& 
$0.19 \pm 0.21$& 
$0.21 \pm 0.35$ \\
$F_T(\%)$&
$1.9 \pm 2.1$&
$2.0 \pm 2.5$&
$2.1 \pm 2.2$&
$1.8 \pm 3.0$ \\
$\phi_{p_T}$(MeV/c)&
$5.65 \pm 6.02$&
$6.03 \pm 7.28$&
$6.11 \pm 6.63$&
$5.47 \pm 9.16$ \\
\end{tabular}
\end{table}

\vspace{1cm}
\begin{table} 
\label{tab:etFluc}
\caption{Fluctuation quantities for the $M_{e_T}$ analysis.}
\begin{tabular}[]{c|cccc}
Centrality&
$0-5\%$&
$0-10\%$&
$10-20\%$&
$20-30\%$ \\ \hline
$\omega_{(T,data)}$(\%)&
$7.32 \pm 0.07$&
$7.84 \pm 0.08$&
$9.58 \pm 0.17$&
$11.8 \pm 0.26$ \\
$d(\%)$&
$0.30 \pm 0.09$& 
$0.37 \pm 0.12$& 
$0.38 \pm 0.20$& 
$0.40 \pm 0.32$ \\
$F_T(\%)$&
$4.3 \pm 1.3$&
$5.0 \pm 1.6$&
$4.2 \pm 2.2$&
$3.5 \pm 2.8$ \\
$\phi_{e_T}$(MeV)&
$11.5 \pm 3.59$&
$13.6 \pm 4.23$&
$11.1 \pm 5.75$&
$9.28 \pm 7.34$ \\
\end{tabular}
\end{table}

\end{multicols}    


\begin{references}
\bibitem[\dagger]{non-par} Not a participating Institution (author is 
an individual participant).
\end{references}
\end{document}